\documentclass[prl,twocolumn,showpacs]{revtex4-1}
\usepackage{graphicx}
\usepackage{bm}
\usepackage{amsmath}
\begin{document}

\title{Looking into the collapse of quantum states with entangled photons}
\author{M. G. M. Moreno and Fernando Parisio}
\email{corresponding author: parisio@df.ufpe.br} 
\address{Departamento de F\'{\i}sica, Universidade Federal de Pernambuco, 50670-901,
Recife, Pernambuco, Brazil}


\begin{abstract}
We propose a scheme to investigate the time scale of the wave-function collapse by using polarization-entangled photon pairs. The setup is similar to those employed to investigate quantum correlations, but in the present case, synchronization is essential at all stages. We find that it is possible to discriminate between the scenarios of instantaneous collapse and finite-time reduction via a large number of double measurements of polarization. The quantities to be recorded would present distinct behaviors in each scenario, the deviations being small but distinguishable from pure statistical fluctuations.
\end{abstract}
\maketitle

\section{Introduction}
%
The wave function and its collapse remain in controversial positions in the general framework of quantum theory. Nevertheless, for long periods in the development of wave mechanics these issues were put aside by most of users of the quantum formalism as, perhaps, an underlying discomfort. One of the reasons for this is the fact that there were plenty of more direct questions to be coped with regarding, e. g., atomic and particle physics. 
In the last few decades, however, experiments reached a remarkable sophistication and  textbook illustrations became feasible in the laboratory. This allowed for objective discussions on, until then, purely academic matters, as for example in the experimental tests \cite{paris} of Bell's inequalities \cite{bell}.  

Since then, part of the focus started to migrate from operational aspects to more foundational ones. This ongoing move is so important that,  justifiably, has been termed {\it the second quantum revolution} \cite{speakable}.
Examples of this process are the debate on the ``reality" of quantum states, which received special attention in the last year \cite{PBR} (see also \cite{schlosshauer2}), and the many facets of the measurement problem, in particular, the collapse of the state vector \cite{gihardi,schlosshauer1,zurek, weinberg, parisio}. These two topics are intimately related since there is no collapse problem in the epistemic view, where a state is regarded as the experimenter's information on some aspects of reality. Particularly, in the statistical interpretation \cite{ballentine}, where the basic entity is an ensemble (nothing being said about single particles), the decoherence program \cite{zurek} alone seems to solve the remaining puzzle, namely, the lack of superposition states in the macroscopic world. However, if one admits that the quantum state of a single object has a physical reality, the ontic view, then the collapse problem persists. In such a case it is hard to accept that any kind of instantaneous evolution can happen. In this work we take this observation earnestly, and argue that if the state vector is of ontological nature, then the collapse should not be instantaneous. 

In what follows we show that it is possible to check this hypothesis experimentally via a large number of synchronized polarization measurements of two correlated photons. It is worth to mention that finite-time collapse has been considered before in  different circumstances, e. g., in the search for stochastic terms which, added to the Shcroedinger equation, produce a reduction dynamics consistent with Born's rule \cite{dynamics}.

\section{Finite-time reduction}
In a scenario of non-instantaneous collapse the measurement postulate must be recast in some way.  A recent proposal \cite{parisio} that we will adopt here, with some modifications of nomenclature, reads :

(I) {\it Measurement duration and ``hits" }: We take into account the fact that any actual measurement has a duration, that we denote by $\Delta t$ and, most importantly, we assume that the collapse is a process initiated by a random hit (we borrow this terminology from \cite{gihardi} in a distinct context) occurring at $t^{(h)}$, which is taken as a stochastic variable obeying some probability distribution defined in the window $[t_0,t_0+\Delta t]$. For $t<t^{(h)}$, the system remains uncoupled to external degrees of freedom. The exact nature of the distribution and the duration  $\Delta t$ depend on the specific system and measurement method, as will be exemplified later.

(II) {\it Finite-time collapse}: The quantum state takes a short time $\delta t$, starting from $t^{(h)}$, to complete the reduction. We do not make any specific statements about the non-unitary time evolution in the interval $[t^{(h)},t^{(h)}+\delta t]$. But we do assume that during the reduction the state of the system is still contained in a ket belonging to the enlarged Hilbert space ${\cal E}_T={\cal E}\otimes {\cal E}_X$, where ${\cal E}$ concerns the system of interest and ${\cal E}_X$ refer to all relevant degrees of freedom that couple with it.

Regarding (I) we remark that, in the case of photodetection, {\it we associate the probability distribution for the occurrence of a hit with the temporal intensity profile of the photon that reach the detector}. Also, it is very important to realize that $t^{(h)}$ must not be confused with the moment when the pointer comes to a definite position, that is, when the avalanche photodiode (APD) delivers a macroscopic current. This time scale has been studied in different perspectives \cite{squires}. It is an essential part of our hypothesis that this macroscopic phenomenon is preceded by a microscopic event that triggers the collapse of the state ket at $t=t^{(h)}$. 

\section{Two correlated photons}
The system we address here is composed of two spatially separated photons which are simultaneously generated, and led to the entangled polarization state 
\begin{eqnarray}
\label{state1}
\nonumber
| \Psi_0 \rangle =\alpha | +\rangle_L \otimes |- \rangle_R +\beta | -\rangle_L \otimes |+\rangle_R  \\
\equiv \alpha | +- \rangle +\beta | -+\rangle\;,
\end{eqnarray}
where $\alpha \ne 0$ and $\beta \ne 0$, and the subscripts $L$ and $R$ refer to the photons sent to the ``left" and ``right" detectors, respectively (Fig. 1). The generation can be achieved with a non-linear crystal via a parametric downconversion process, which, in general, gives synchronization and may also produce entanglement in polarization \cite{dc1,dc2}. Since the photons follow distinct optical paths, a delay between them is likely to be introduced. While this is not critically relevant in evaluating violations of Bell's inequalities, it may hinder the phenomenon we intend to investigate. Up to this point, the synchronization of the pair can be restored with the help of a Hong-Ou-Mandel (HOM) apparatus \cite{hom} and delay lines coupled with translation stages. From this point to the detectors the synchronization is technically non-trivial, but can be handled, in principle. For a recent proposal of a scheme to measure ultra-short delays see \cite{synchron}. As the photons reach the detectors their polarizations are measured with the filters set in the {\it same} direction, for which $\{|+ \rangle,|-\rangle\}$ are eigenstates.

To be more realistic, we assume that the two wave packets attain the detectors simultaneously (in the laboratory frame), except by a delay $\cal{T}$ that eventually persists (Fig. 2).
We stress that our model encompasses the expected situation in which the residual delay is typically much larger than $\delta t$. Note carefully that the procedure ensures that the centroids of each wave packet will reach the detectors approximately at the same time, and not that the (unpredictable) hits themselves will be simultaneous. 
Note also that the spatio-temporal profile to be considered is not that of the generated photons, but rather of the photons just before detection (with the spreading and deformation taken into account). Finally, the quantities we suggest to be measured are subtle statistical deviations, so we need a robust sampling. This demand naturally leads us to consider that a pulsed laser with a high repetition rate is employed as the primary source of photons.
\begin{figure}
\label{figure1}
\includegraphics[width=5cm,angle=0]{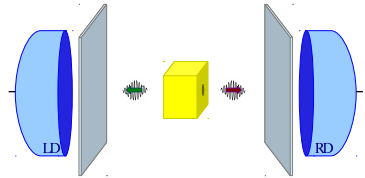}
\caption{(color online) Pictorial representation of the experimental setup. The box contains a non-linear crystal that generates pairs of synchronized photons via PDC. The crystal is pumped by a pulsed laser (not shown).}
\end{figure}

From basic quantum mechanics we immediately infer the statistical distribution resulting from a series of $N$ coincidence polarization  measurements on state (\ref{state1}), see table \ref{table1}.
\begin{table}[h]
\centering
\begin{tabular}{| c | c | c |}
\hline
 Result & Frequency \\ \hline
 Left $+$ , Right $-$ &  $|\alpha|^2$ \\ \hline
 Left $-$ , Right $+$ & $1-|\alpha|^2$\\ \hline
\end{tabular}
\caption{Outputs and relative frequencies of two sequential polarization measurements in the same direction according to quantum mechanics. The second measurement plays no role since, after the first one, the state collapses instantly.}
\label{table1}
\end{table}
Fluctuations with magnitude $\Delta {\cal N} \sim \sqrt{N}/2$ naturally show up for any finite number of repetitions. Since in any standard interpretation of quantum mechanics the collapse is assumed to be instantaneous, the second measurement of polarization (in the same direction) would not play any role. 

Now, we address the same question, this time considering the possibility of non-instantaneous collapse. In this case we must analyze the development of the events more carefully. We start by assuming that the intensity profile of the electromagnetic field associated to the photons is already characterized. 
We denote the distributions for a hit in the left and right detectors by $f_L(t)$ and $f_R(t)$, respectively, and, for definiteness, we assume the left photon to be delayed with respect to the right one. Apart from this we consider the two packets as having the same shape, that is
\begin{equation}
\label{delay}
f_R (t)=f_L(t-\cal{T}) \;.
\end{equation}
\begin{figure}
\label{figure2}
\includegraphics[width=5cm,angle=0]{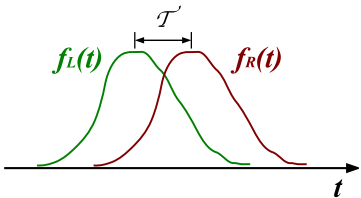}
\caption{(color online) Probability densities for the occurrence of a hit at the left and right photons. The residual delay, due to imperfections in the synchronization process is denoted by $\cal{T}$.}
\end{figure}
In this scenario two distinct situations may happen. If, as we suppose here, $\delta t$ is smaller than any other time scale in the problem, then, with high probability, when the second hit takes place, the reduction due to the first one is already completed. For these realizations we obtain exactly the results shown in Table \ref{table1}. However, in a small number of non-trivial events, according to our hypothesis, the second hit catches the state ket while it is still collapsing.
Setting $y=t_L^{(h)}-t_R^{(h)}$, where $t_L^{(h)}$ [$t_R^{(h)}$] is the time when a hit occurs in the left (right) detector, the probability $P(|y|<\delta t)\equiv P_<$ for the occurrence of both hits in a time interval shorter than $\delta t$ is
\begin{equation}
\label{prob}
P_<=\int_{-\Delta t}^{+\Delta t} \int_{t_2=t_1-\delta t}^{t_2=t_1+\delta t} f_L(t_1)\,f_R(t_2) \,{\rm d}t_1 \, {\rm d}t_2\;.
\end{equation}
The probability density for the relative variable $y$ is given by $p(y)={\rm d}P/{\rm d}(\delta t)|_y$.

As soon as the first hit happens, no matter in what detector (the two filters are parallel), the state starts to collapse following one of the two kinematic routes
\begin{equation}
\label{route1a}
| \Psi_1(t) \rangle =a_1(t) | +- \rangle\otimes |\Phi_{+-}^{(1)}\rangle +b_1(t) | -+\rangle \otimes |\Phi_{-+}^{(1)}\rangle \;,
\end{equation}
\begin{equation}
\label{route2a}
| \Psi_2(t) \rangle =a_2(t) | +- \rangle\otimes |\Phi_{+-}^{(2)}\rangle +b_2(t) | -+\rangle \otimes |\Phi_{-+}^{(2)}\rangle \;,
\end{equation}
with boundary conditions
$a_k[t^{(h)}]=\alpha, a_k[t^{(h)}+\delta t]=\delta_{1,k}, b_k[t^{(h)}]=\beta, b_k[t^{(h)}+\delta t]=\delta_{2,k}$, where $k=1,2$. We dispensed with the subscripts $L$ and $R$ and denote the instant when the first hit happens by $t^{(h)}$. The kets $|\Phi\rangle$ correspond to the microscopic states of the degrees of freedom that couple to the system. 
To be consistent with Born's rule we assume that route (\ref{route1a}) happens with relative frequency $|\alpha|^2$ and the second route, Eq. (\ref{route2a}), with relative frequency $|\beta|^2=1-|\alpha|^2$. We use the terminology ``kinematic route" because we are not providing, or trying to provide, the dynamical equations that are satisfied by $a_k(t)$ and $b_k(t)$ during the reduction. Instead, we only use the fact that at the end of the process Born's postulate must be verified.  We also remark that by excluding terms proportional to $ | ++ \rangle$ and $ | -- \rangle$ in the intermediate states (\ref{route1a}) and (\ref{route2a}), with vanishing coefficients for $t=t^{(h)}$ and $t=t^{(h)}+\delta t$, we assume strict angular momentum conservation during the whole process.

Let us then consider the rare event of a second hit happening between $t$ and $t +{\rm d}t$ with $t^{(h)}<t<t^{(h)}+\delta t$. In this situation, if the state is catch collapsing via route (\ref{route1a}), the outcomes after the second reduction is completed are
$| +- \rangle |\Phi_{+-}^{1f}\rangle$ with probability $|a_1(t)|^2$ and
$| -+\rangle  |\Phi_{-+}^{1f}\rangle$ with probability $|b_1(t)|^2$.
If the state is collapsing through route (\ref{route2a}), the possible results are
$| +- \rangle |\Phi_{+-}^{2f}\rangle$ with probability $|a_2(t)|^2$ and
$| -+\rangle  |\Phi_{-+}^{2f}\rangle$ with probability $|b_2(t)|^2$. The final states $| \Phi^f \rangle$ need not be macroscopic pointers at this stage. Rather, we assume that they contain the state of the apparatus which, after some extra time, will describe a definite macroscopic pointer position. We, thus, have a composite von Neumann chain.
From the previous reasoning, the probability of getting $+-$ between $t$ and $t +{\rm d}t$ is
proportional to
\begin{equation}
\left[|\alpha|^2|a_1(t)|^2+ (1-|\alpha|^2)|a_2(t)|^2\right]{\rm d}t\;.
\end{equation}
The probability of obtaining $+-$ for any $t$, satisfying $t^{(h)}<t<t^{(h)}+\delta t$, is given by the integral
\begin{eqnarray}
\nonumber
\frac{1}{P_<}\int_{-\delta t}^{\delta t}[|\alpha|^2|a_1(y)|^2+ (1-|\alpha|^2)|a_2(y)|^2]p(y){\rm d}y\\
=\frac{|\alpha|^2 \Gamma+\Lambda}{P_<}\equiv
P(+-|\;|y|<\delta t)\;,
\end{eqnarray}
where
\begin{equation}
\Lambda=\int_{-\delta t}^{\delta t}|a_2(y)|^2p(y){\rm d}y \;,\Gamma=\int_{-\delta t}^{\delta t}|a_1(y)|^2p(y){\rm d}y - \Lambda\;.
\end{equation}
We can write the unconditional probability of getting the result $+-$ as
\begin{eqnarray}
\nonumber
P(+-)=(1-P_<)|\alpha|^2+P_<\,P(+-|\;|y|<\delta t)\\
=|\alpha|^2+[|\alpha|^2(\Gamma-P_<)+\Lambda]\;.
\end{eqnarray}
The above result can be further simplified if we realize the existence of a discrete left-right symmetry, that must hold. Had we considered an initial state with $\alpha \leftrightarrow \beta$, we would get
$P(+-) \leftrightarrow P(-+)$, where $P(-+)=1-P(+-)$. That is
$P_{(|\alpha|^2 \leftrightarrow 1- |\alpha|^2 )}(+-)=P(-+)=1-P(+-)$,
which leads to the constraint $\Gamma=1-2\Lambda$. Finally we obtain
\begin{equation}
P(+-)
=|\alpha|^2+(1-2|\alpha|^2)\Lambda \;.
\end{equation}
Therefore, if the collapse is not instantaneous, within our hypothesis, the outcomes of two well synchronized polarization measurements should be characterized by Table \ref{table2}.
\begin{table}[h]
\centering
\begin{tabular}{| c | c | c |}
\hline
 Result & Frequency \\ \hline
 Left $+$ , Right $-$ &  $|\alpha|^2+(1-2|\alpha|^2)\Lambda $ \\ \hline
 Left $-$ , Right $+$ & $1-|\alpha|^2-(1-2|\alpha|^2)\Lambda $\\ \hline
\end{tabular}
\caption{Outputs and relative frequencies of two sequential, accurately synchronized, polarization measurements in a scenario of finite-time reduction.}
\label{table2}
\end{table}

The above table might give the impression that we are suggesting a correction to Born's postulate. This is not the case, since the postulate refers to the likelihood of each possible result of a single measurement. If collapse is indeed instantaneous, a second measurement of the same observable would be innocuous. What we have just shown is that, if the collapse takes a finite time, then a close consideration of Born's rule leads to the probabilities in table \ref{table2}. Once a sufficiently large number $N$ of repetitions is made, the numeric difference between the results $+-$ and $-+$, according to table \ref{table1} is $\Delta N_{I}=(2|\alpha|^2-1)N$,
while the same quantity, according to table \ref{table2}, is $\Delta N_{II}=(2|\alpha|^2-1)(1-2\Lambda)N$. Thus, the deviation between the two scenarios is given by
\begin{equation}
\label{difference}
\Delta N= \Delta N_{I}-\Delta N_{II}=2\Lambda(2|\alpha|^2-1)N\;.
\end{equation}
It is clear that a maximally entangled state, with $|\alpha|=1/\sqrt{2}$, would not reveal a potentially non-vanishing result. The initial state (\ref{state1}) have to be unbalanced.  As we will see next, $\Lambda$ is typically a very small number and the difference
(\ref{difference}) is subtle. The immediate question that arises is, being $\Delta N/N$ small, can we safely distinguish it from pure statistical fluctuations ($\Delta {\cal N}$) that 
surely occur in an actual experiment? Fortunately, the deviation in the worse scenario, where $\Delta {\cal N}_{I} \sim - \sqrt{N}/2$ and $\Delta {\cal N}_{II}\sim  \sqrt{N}/2$ tend to minimize $\Delta N$ for $|\alpha|> 1/\sqrt{2}$,  is $\Delta {\cal N} = \sqrt{N}$, while $\Delta N \sim N$, so that for a sufficiently large number of realizations one can reach a ratio $\Delta N/\Delta {\cal N} $ as large as needed. In fact, $\Delta N/\Delta {\cal N} = 2\Lambda(2|\alpha|^2-1)\sqrt{N}$, and the number of realizations must satisfy
\begin{equation}
N>K^2 \times [2\Lambda(2|\alpha|^2-1)]^{-2}\;,
\end{equation}
for a statistical significance of $K$ standard deviations ($\Delta N> K \Delta {\cal N}$). 

Once the general framework is set, let us examine a specific example. Suppose the source of light is a pulsed titanium-sapphire laser whose temporal profile of intensity reads 
\begin{equation}
\label{profile}
f(t)=\frac{1}{2\sigma_t }\text{sech}^2\left(\frac{t}{\sigma_t} \right)\;,
\end{equation}
where the pulse width $\sigma_t$ provides the coherence time. By using (\ref{delay}) and (\ref{prob}) we get
\begin{eqnarray}
\label{prob2}
\nonumber
P_<=\frac{1}{4} \sum_{n=0,1} (-1)^n\left\{ \text{csch}^2(A_n)\ln\left[ \frac{\cosh(A_n+\Delta t/\sigma_t)}{\cosh(A_n-\Delta t/\sigma_t)}\right]\right.\\
\left. -2\coth(A_n)\tanh(\Delta t/\sigma_t)\right\}\;,\hspace{1cm}
\end{eqnarray}
with $A_n=[{\cal T}+(-1)^{n+1}\delta t]/\sigma_t$. Assuming that $\delta t$ and ${\cal T}$ are much smaller than the coherence time, the above result simplifies to
\begin{equation}
P_<\approx \frac{1}{2}\tanh\left( \frac{\Delta t}{\sigma_t}\right)\left[  \tanh\left( \frac{{\cal T}+\delta t}{\sigma_t}\right)-\tanh\left( \frac{{\cal T}-\delta t}{\sigma_t}\right)\right]\;,
\end{equation}
leading to the probability distribution
$2\sigma_t p(y)\approx \tanh\left( \Delta t/\sigma_t\right) \text{sech}^2\left[({\cal T}+y)/\sigma_t\right]$. 
A typical repetition rate of a pulsed laser is 100 MHz, however, this is not the frequency at which the correlated pairs are detected in coincidence. We assume that this rate is diminished in three orders of magnitude, giving one detection in coincidence per 10 $\mu$s, on average. Furthermore, the residual delay is of the order of $\delta L/c$, where $\delta L$ is the step of the translation stage in the delay line and $c$ is the velocity of light. Usually $\delta L \approx 1\,\mu$m, so ${\cal T} \approx 3.3$ fs. The duration $\Delta t$ of each measurement is set to ensure that the detected photons belong to the same pair. We can safely consider the window of coincidence to be $\Delta t \approx 1$ ns. Finally, for the sake of illustration let us assume that the final shape of the wave packets at detection is still given by Eq. (\ref{profile}), with a relatively large spreading of $\sigma_t \approx 1$ ps (the coherence time soon after the generation is, say, 200 fs).

Consider that we intend to investigate the compatibility of experimental data with a collapse in the range of $\delta t \sim 0.1$ fs. This would lead to $P_< \approx 10^{-4}$, corresponding to 10 non-trivial detections per second. 
Of course, in order to get numbers we must assume some functional form for $|a_2(t)|$ before calculating $\Lambda$. The quantitative results weakly depend on this choice, but the qualitative features remain unchanged.
By choosing an exponential decay for $|a_2(t)|$, satisfying the appropriate boundary conditions, we get $\Lambda \approx 2.0 \times 10^{-4}$ \cite{comment}.  Suppose that, with $\alpha=\sqrt{3}/2$, we obtain a reliable statistics characterized by $\Delta N/\Delta {\cal N} \sim 6$ (six standard deviations), corresponding to a 12 hour long experiment ($N \approx   10^9$ realizations). This result alone would be a strong evidence for finite-time collapse. Of course, further experimentation would be necessary, varying ${\cal T}$, $\alpha$, and the orientation of the filters, to investigate the actual time dependence of $a_2(t)$. It would be especially important to repeat the same procedure with the filters set in orthogonal directions for, in this situation, there should be no measurable difference between the two scenarios for any pair $\alpha$, $\beta$. Conversely, if in the original experiment we obtain $\Delta N/\Delta {\cal N}\sim1$, then 0.1 fs would be an upper bound for an exponential reduction in the system studied. Lastly, the above estimates would not change appreciably for ${\cal T}=0$, showing that the scheme is robust for delays of order of femtoseconds.

\section{Conclusions}
It might be considered insufficient to assert that if the state is $\psi$-ontic, in the sense adopted in the literature \cite{comment_b}, then the wave function is a real thing \cite{comment_c}. A possibly reasonable extra requirement would be that the microscopic collapse should not be instantaneous . We stress that this has nothing to do with the condition of locality, since the collapse $(|a\rangle |b \rangle-|b\rangle|a\rangle)/\sqrt{2} \overset{\delta t} \longrightarrow |a\rangle |b \rangle$ is, in general, non-local, provided that the subsystems are sufficiently far apart.
By employing minimal statements (of kinematic nature) about this finite-time collapse and assuming that Born's rule remains valid during the non-unitary evolution, we claim that it is possible to probe the collapse duration experimentally in the scale of subfemtoseconds. Although decoherence is not a logical necessity of our model, the previous results are not incompatible with it in any obvious way. In fact, it has been suggested by Schlosshauer \cite{schlosshauer1} that a combination of dynamical localization models and the effects of environment is a promising strategy to approach the collapse problem.

An important point is the lack of covariance of our results, which is not an exception in dealing with entanglement. If we admit that the two hits are not causally related, their ordering may be swapped for some inertial frame. Alternatively, if one assumes that the hits are causally connected it is necessary to admit the existence of an ``ether" in which the spatially separated subsystems exchange information via a supraluminal signaling. A stringent lower bound for the velocity of this signal has been placed \cite{spooky}, $v_{signal}\sim 10^4c$ for a continuous set of referentials whose relative velocity with respect to earth is as large as $0.1c$ . Although a detailed discussion of this point is outside the scope of the present work, we believe that the dynamics of collapse, and its time scale, deserves investigation in either case.
\begin{acknowledgements}
The authors warmly thank Katiuscia Cassemiro, Daniel Felinto, and A. M. S. Mac\^edo for many relevant discussions on this work.
F.P. thanks the comments by the members of the quantum optics and quantum information group at the Universidade Federal Fluminense. Funding from CNPq, CAPES, and FACEPE (APQ-1415-1.05/10) is acknowledged.
\end{acknowledgements}


\begin{thebibliography}{99}
\bibitem{comment} For a drastically distinct time dependence, a linear decay of $a_2(t)$, we get 
$\Lambda \approx 1.0 \times 10^{-4}$, that would not change the order of magnitude of our results.
\bibitem{comment_b} The state is $\psi$-ontic if the support in the space of ``hidden" variables $\lambda$  of an arbitrary state 
$|\psi \rangle$ does not overlap the support of any other state  $|\phi \rangle$. That is, from the distribution of states $\lambda$ 
one can unambiguously determine the state ket of the system.
\end{thebibliography}
\end{document}